# Non-equilibrium magnetism in dual spin valves

A. Aziz[1], O. P. Wessely[2], M. Ali[3], D.M. Edwards[2], C.H. Marrows[3], B.J. Hickey[3], M.G. Blamire[1]

[1] Department of Materials Science, Cambridge University, Pembroke Street, Cambridge CB2 3QZ, UK

[2] Department of Mathematics, Imperial College, London SW7 2BZ, UK

[3] School of Physics and Astronomy, University of Leeds, Leeds LS2 9JT, UK

**The field of spin electronics (spintronics) was initiated by the discovery of giant magnetoresistance (GMR) for which Fert[1] and Grünberg[2] were awarded the 2007 Nobel Prize for Physics. GMR arises from differential scattering of the majority and minority spin electrons by a ferromagnet (FM) so that the resistance when the FM layers separated by non-magnetic (NM) spacers are aligned by an applied field is different to when they are antiparallel. In 1996 Slonczewski[3] and Berger[4] predicted that a large spin-polarised current could transfer spin-angular momentum and so exert a spin transfer torque (STT) sufficient to switch thin FM layers between stable magnetisation states[5] and, for even higher current densities, drive continuous precession which emits microwaves[6]. Thus, while GMR is a purely passive phenomenon which ultimately depends on the intrinsic band structure of the FM, STT adds an active element to spintronics by which the direction of the magnetisation may be manipulated. Here we show that highly non-equilibrium spin injection can modify the scattering asymmetry and, by extension, the intrinsic magnetism of a FM. This phenomenon is completely different to STT and provides a third ingredient which should further expand the range of opportunities for the application of spintronics.**

The basic theories of GMR[7,8] are linear so that the magnitude of the MR is independent of current density. Conventional applications of the effect, for example in the FM/NM/FM "spin-valve" (SV) used for reading data from hard discs, use relatively low current densities. In contrast, STT switching current densities are high (typically $\sim 10^7$ Acm$^{-2}$) resulting in substantial spin-accumulation in the NM spacer layers. Much less attention has been paid to spin-accumulation in FMs; here we show that in dual spin-valves (DSV) which are configured so that the outer FM layers are antiparallel to enable extreme spin-accumulation in ultra-thin middle FM layer, the intrinsic properties of this FM can be modified so that a current-dependent non-equilibrium magnetic state is created. The signature of this non-equilibrium magnetism is a *current-dependent* MR which is fundamentally different to STT switching.

In our DSV structures the active layers consist of a sputter-deposited $Co_{90}Fe_{10}$(6nm) /Cu(4nm) /Py(x) /Cu(y) /$Co_{90}Fe_{10}$ (6nm) /IrMn (10nm) stack (different samples are henceforth labelled as DSV(x,y)); Py is permalloy ($Ni_{80}Fe_{20}$). Thick Cu layers above and below this stack were used for the bottom and the top electrical contacts and a 3-D gallium focused ion beam milling technique was used for fabricating current perpendicular to plane nanopillar devices (Fig. 1a); fuller fabrication details are can be found in reference[9]. The measurements presented here are performed using asymmetric DSVs with y = 2 nm, but symmetrical devices DSV(x,4)) behave similarly; single SVs (DSV(0,2)) have also been measured. All the resistance measurements have been performed at room temperature using a lock-in technique with frequency 77Hz and $I_{ac}$ = 100 µA. Positive $I_{dc}$ corresponds to the electrons flowing from top to bottom i.e. from IrMn/CoFe to CoFe layer.

Figure 1b shows the full MR loop of a DSV(2,2) sample for $I_{dc}$ = 0. The patterned devices are too small for the magnetic configurations to be measured directly and so careful finite-element magnetic simulations have been performed (see Supplementary Information) to identify the magnetic configurations corresponding to the various plateaux in the loop. At the starting magnetic field (*H*) of +150 mT all the magnetic layers are parallel and so the MR is at its lowest value[7]. When *H* is decreased to about 50 mT, the soft Py layer reverses under the influence of the magnetostatic fields of the parallel-aligned top and the bottom CoFe layers. In essence this new state can be viewed as two anti-parallel spin valves in series and so there is a substantial increase in resistance. Further reduction in *H* results in the reversal of one of the CoFe layers, evident as a small drop in the MR close to zero field, to form a series combination of a parallel and an antiparallel SV. The second CoFe layer reverses at about -70mT, resulting in a fully parallel state. Although the samples contained an IrMn pinning layer only a small bias was observable as an asymmetry in the maximal switching fields, but this enables confirmation of the magnetic configurations.

To access other states in which the CoFe layers are antiparallel, minor loop measurements were performed in which the system was first saturated at +150 mT and then loops were measured by sweeping *H* between ±32 mT. Fig. 1c1 shows the minor loop for $I_{dc}$=0 mA: the initial path from +32 mT to –32 mT is shown in black and here the MR decrease is due to the switching of the CoFe layer as in the major loop. In, contrast, within the stable minor loop it is the Py layer which reverses close to zero field and the actual change in resistance is minimal (confirmed from the simulations shown in the Supplementary Information).

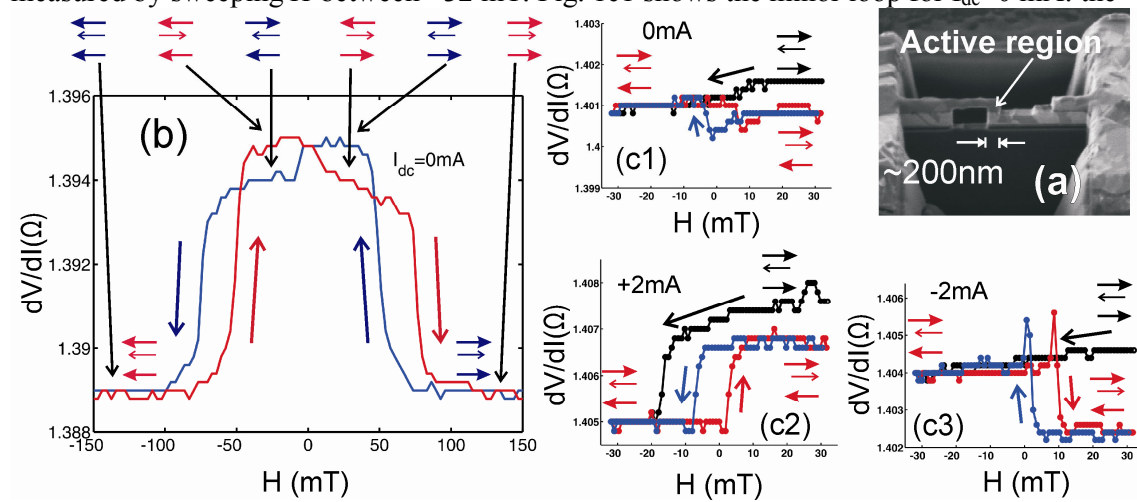

Figure 1| **a**, Micrograph of a device: the DSV is within the active region where the current flows perpendicular to the plane. **b** Low current full resistance vs field loop of a 120 × 190 nm$^2$ DSV(2,2) device; red and blue lines correspond respectively to increasing and decreasing field sweeps and the arrows at the top of the graphs represent the magnetic states of the CoFe (top), Py(middle) and CoFe (bottom). **c** Minor MR loops measured at 0 and ±2mA: black curve is initial sweep from +150 mT to –32 mT; blue and red curves sweep between ±32 mT as above.

The behaviour which is the subject of this Letter is the dramatic change in minor loop shape which occurs when large currents are applied (Fig. 1c2, 1c3). Unlike changes associated with STT-switching the primary changes are in the *magnitude* of the resistance corresponding to each magnetic state rather than in the transition fields between them. The sign of the resistance change on crossing zero field is dependent on the current direction and the magnitude of the change is much larger for $I_{dc}= \pm 2$ mA than for zero current. Interestingly, the same current dependence has also been observed for DSV(1,2) and DSV(1,4) samples.

Figure 2a shows in detail the current-dependence of the stable minor MR loops for DSV(1,2); the minor loops at ± 1 mA are strikingly similar to those for DSV(2,2) at ± 2 mA (Fig. 1c). To compare devices with different Py thicknesses we plot the field-induced change in resistance-area product ($A\Delta R$) vs current density for different Py thicknesses (Fig. 2b); the current-dependence of $A\Delta R$ decreases rapidly with Py thickness.

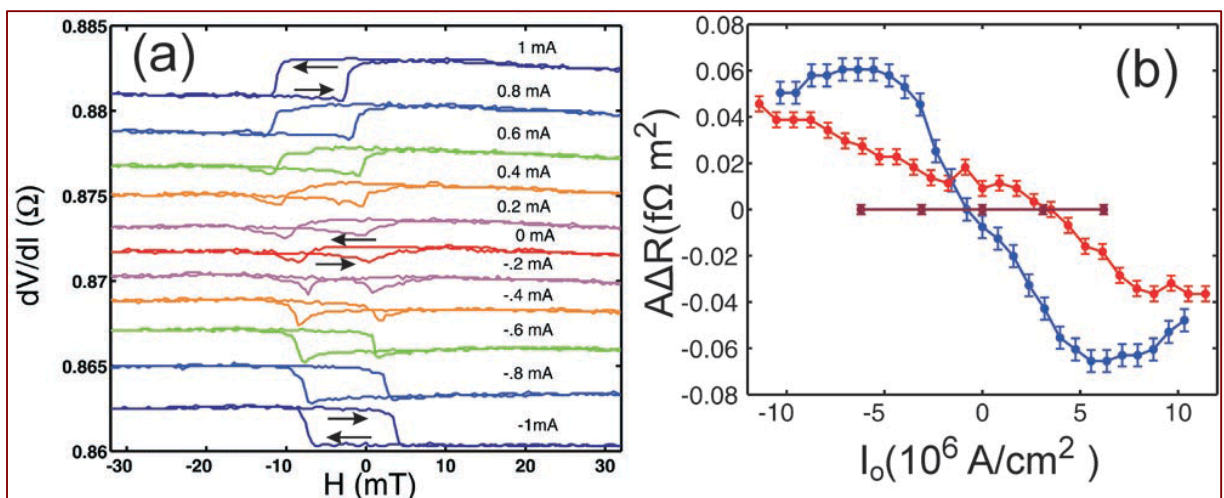

Figure 2| **a** A series of minor differential resistance vs magnetic field for a DSV(1,2) device for dc currents varying from -1mA to +1mA. **b** The change in resistance area product vs current density for DSV(1,2) (blue), DSV(2,2) (red) and DSV(8,2) (magenta).

To reveal the current-induced reversal of the MR more explicitly, 3-D maps of the minor loop MR for DSV(1,2) are drawn in Figs. 3a&b, showing the differential resistance as a function of current and magnetic field for decreasing and increasing $H$ respectively. Both diagrams show that high resistance states transform gradually to low resistance states when current is increased in the positive (negative) direction in negative (positive) $H$. However the change in resistance is abrupt (and hysteretic) along $H$ demonstrating that only the *field* and not the *current* controls the magnetic state. Figure 3c shows the traces of differential resistance versus current extracted from Fig. 3b at ±20mT. An essentially identical curve is also obtained by sweeping current at a constant magnetic field of -20mT as shown in Fig. 3d; this confirms that the change in resistance

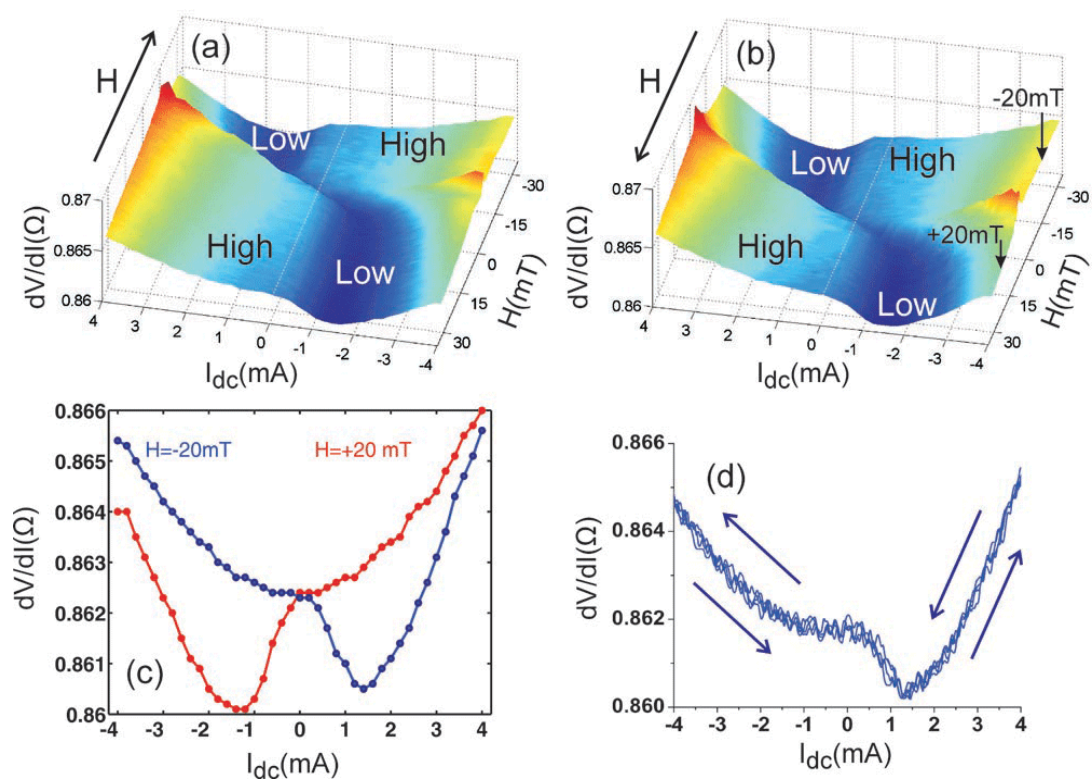

FIG. 3| Measurements of a DSV(1,2) sample: **a** and **b** MR as a function of dc current: magnet field sweep directions as indicated. Low (dark blue) and high (yellow) resistance states are marked. In all cases the parabolic background is due to Joule heating. **c** Differential resistance vs $I_{dc}$ extracted profiles from figure b at ±20mT. **d** Directly measured differential resistance vs $I_{dc}$ at -20 mT.

along the current direction is reversible and non-hysteretic with respect to current.

Our data shows that the MR is essentially independent of current when the outer CoFe layers are parallel or when the Py is omitted (DSV(0,2) single spin-valve) and the pronounced changes occur when these layers are antiparallel (as in the case of the minor loops) and a substantial spin accumulation occurs in the Py. Neither conventional GMR nor STT can explain the observed behaviour although at higher current densities we do see peaks in the differential conductance which are consistent with STT-induced precessional states.

In the Valet and Fert[7] model for GMR, the MR is determined by the bulk and interfacial spin-scattering asymmetries of the FMs; these coefficients are conventionally fixed and depend on the shapes of the majority and minority Fermi surfaces[10] so that the MR is current-independent. Here we are proposing that sufficient spin accumulation results in a current-controlled non-equilibrium magnetic state in which the spin-splitting of the density of states (DoS) in the middle FM layer and hence spin-dependent scattering are current-dependent.

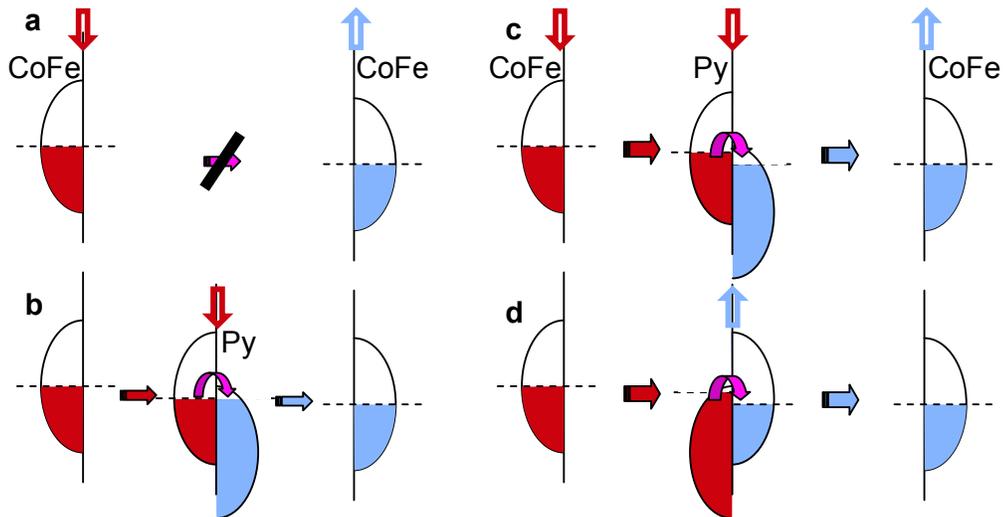

FIG. 4| Schematic diagrams of current flow in DSV devices with half-metallic outer electrodes: **a** absence of current without permalloy layer (Py); **b** spin flip in Py enables current flow; **c**, **d** Large currents result in decreased and increased spin-splitting in Py for alternative Py orientations.

To understand the behaviour more explicitly we focus on the importance of spin-flip processes in determining the overall conductivity of DSV structures. For simplicity we consider the behaviour of a somewhat idealised system in which the outer CoFe layers are replaced by ideal half-metallic ferromagnets. When the Py layer is absent (Fig. 4a) the system becomes a single spin valve and the current in the antiparallel configuration would then be zero as neither spin-up or -down channels would be open. The addition, to form the DSV, of a conventional (partially spin-polarised) ferromagnetic middle layer in which spin-flip processes are possible allows the necessary conversion of the spin current from up to down for current flow (Fig. 4b). The transport currents therefore carry a net up-spin current into the Py and, in the conventional Valet & Fert model, creates a difference ($\Delta\mu$) in the chemical potentials of spin-up and spin-down electrons which is proportional to the current and inversely proportional to the thickness. Using parameters appropriate to our experiment suggests that $\Delta\mu \sim 1$ meV which, in a rigid band picture of ferromagnetism in Py, seems too small to induce direct changes in the scattering asymmetry and hence the MR. What this appears to imply is that large $\Delta\mu$ (at least in comparison with previous spin accumulation experiments) can couple to the exchange splitting of the density states. On this basis, the asymmetry in the response of the MR to current direction seen in Fig. 3c and 3d could be due to suppression of the exchange splitting for one sign of $\Delta\mu$ which effectively weakens the magnetism of the Py, whereas reversing the current and hence $\Delta\mu$ would have to enhance the exchange splitting. Whatever the origin, the DSV behaves as a weak diode which can be reversed by switching the Py moment and so could, in optimised systems, form the basis of logic operations.

1. Baibich, M. N. et al. Giant Magnetoresistance Of (001)Fe/(001) Cr Magnetic Superlattices. *Phys. Rev. Lett.* **61**, 2472-2475 (1988).
2. Binasch, G., Grunberg, P., Saurenbach, F. & Zinn, W. Enhanced Magnetoresistance in Layered Magnetic-Structures with Antiferromagnetic Interlayer Exchange. *Phys. Rev. B* **39**, 4828-4830 (1989).
3. Slonczewski, J. C. Current-driven excitation of magnetic multilayers. *J. Magn. Magn. Mater.* **159**, L1-L7 (1996).
4. Berger, L. Emission of spin waves by a magnetic multilayer traversed by a current. *Phys. Rev. B* **54**, 9353-9358 (1996).

5. Myers, E. B., Ralph, D. C., Katine, J. A., Louie, R. N. & Buhrman, R. A. Current-induced switching of domains in magnetic multilayer devices. *Science* **285**, 867-870 (1999).
6. Kiselev, S. I. et al. Microwave oscillations of a nanomagnet driven by a spin-polarized current. *Nature* **425**, 380-383 (2003).
7. Valet, T. & Fert, A. Theory of the Perpendicular Magnetoresistance in Magnetic Multilayers. *Phys. Rev. B* **48**, 7099-7113 (1993).
8. Levy, P. M. & Zhang, S. F. Resistivity due to domain wall scattering. *Phys. Rev. Lett.* **79**, 5110-5113 (1997).
9. Wu, M. C. et al. Room temperature spin-transfer effect in exchange-biased spin valve nano-pillars fabricated by 3-D focused-ion beam lithography. *Appl. Phys. Lett.* **92**, 142501 (2008).
10. Stiles, M. D. Spin-dependent interface transmission and reflection in magnetic multilayers. *J. Appl. Phys.* **79**, 5805-5810 (1996).
11. Stein, S., Schmitz, R. & Kohlstedt, H. Magneto-tunneling injection device (MAGTID). *Sol. St. Commun.* **117**, 599-603 (2001).
12. Bass, J. & Pratt, W. P. Spin-diffusion lengths in metals and alloys, and spin-flipping at metal/metal interfaces: an experimentalist's critical review. *J Phys: Condens. Matter* **19**, 183201 (2007).
13. Soulen, R. J. et al. Measuring the spin polarization of a metal with a superconducting point contact. *Science* **282**, 85-88 (1998).

Acknowledgements : This work was supported by EPSRC through the Spin@RT consortium.